\documentclass[pra,aps,showpacs,twocolumn,superscriptaddress,nofootinbib,floatfix]{revtex4}
\usepackage{hyperref}
\usepackage{bm}
\usepackage{epsfig}
\usepackage{amssymb}
\usepackage{graphicx}
\usepackage{amsmath}
\usepackage{epstopdf}
\usepackage{times,txfonts}

\newcommand{\ket}[1]{|#1\rangle}
\newcommand{\bra}[1]{\langle#1|}
\begin{document}
\title{No-activation theorem for Gaussian nonclassical correlations by Gaussian operations}

\author{Ladislav Mi\v{s}ta, Jr.}
\affiliation{Department of Optics, Palack\' y University,
17.~listopadu 12,  771~46 Olomouc, Czech Republic}

\author{Daniel McNulty}
\affiliation{Department of Optics, Palack\' y University,
17.~listopadu 12,  771~46 Olomouc, Czech Republic}

\author{Gerardo Adesso}
\affiliation{School of Mathematical Sciences, The University of
Nottingham, University Park, Nottingham NG7 2RD, United Kingdom}

\date{\today}

\begin{abstract}
We study general quantum correlations of continuous variable Gaussian states and their interplay with entanglement. Specifically,
we investigate the existence of a quantum protocol activating all nonclassical correlations between the subsystems of an input bipartite
continuous variable system, into output entanglement between the system and a set of ancillae. For input Gaussian states, we prove that such an
activation protocol cannot be accomplished with Gaussian operations, as the latter are unable to create any output entanglement from an initial separable
yet nonclassical state in a worst-case scenario. We then construct a faithful non-Gaussian activation protocol, encompassing infinite-dimensional
generalizations of controlled-NOT gates to generate entanglement between system and ancillae, in direct analogy with the finite-dimensional case. We finally
calculate the negativity of quantumness, an operational measure of nonclassical correlations defined in terms of the performance of the activation protocol,
for relevant classes of two-mode Gaussian states.
\end{abstract}
\pacs{03.65.Ud, 03.67.Ac, 42.50.Dv}

\maketitle

\section{Introduction}

Quantum correlations in composite systems transcend entanglement \cite{modirev}.
A bipartite quantum state $\rho_{AB}$ can be defined as nonclassical or
nonclassically correlated if it cannot be expressed as a convex
mixture of local basis states of subsystems $A$ and $B$ \cite{nolocal}. Consequently, all
inseparable (entangled) states as well as the majority of separable states are  nonclassical.

General nonclassical correlations, however, can be mapped to entanglement in a very precise sense, which provides an insightful framework for their characterization and operational interpretation. Specifically, it was proven in \cite{Piani_11,streltsov,PianiAdesso} and very recently experimentally observed in \cite{sciarrino} that all nonclassical states of a finite-dimensional system can be
turned into states with distillable entanglement between the system and a set of ancillae by an {\it activation
protocol}. Focusing on a bipartite setting, the protocol runs as follows. The subsystems $A$ and $B$ are first subject to arbitrary local unitary
transformations $U_{A,B}$; then, each system $j=A,B$ interacts via a controlled-NOT (CNOT) operation
$U_{jj'}^{CNOT}$ (i.e.~a so-called premeasurement interaction) with an auxiliary system $j'$, $j=A,B$, initialized
in a pure state $|0\rangle_{j'}$.
The activation protocol then possesses two key properties: i) for all classical states
$\rho_{AB}$ at the input of the protocol, there exist local
unitaries $U_{A,B}$ for which the output state $\rho_{ABA'B'}$ is
separable across the $AB|A'B'$ splitting, and ii) for all
nonclassical states $\rho_{AB}$ and for all local
unitaries, the output state is entangled across the
$AB|A'B'$ splitting.

Let us stress that both criteria i) and ii) must be met by any scheme in order
to be a valid activation protocol. In particular, they allow us to define faithful measures of nonclassical correlations for the input state  $\rho_{AB}$ in terms of the output  $AB|A'B'$ entanglement, minimized over $U_{A,B}$. One such measure, when the output entanglement is quantified by the negativity \cite{Vidal_02}, has been termed negativity of quantumness \cite{Piani_11,Takafumi_13}, and has been experimentally investigated in \cite{isabela,sciarrino}

In this paper we study activation of nonclassical correlations in
multimode bipartite Gaussian states  $\rho_{AB}$ of
continuous variable systems \cite{ourreview}. Nonclassical
correlations of Gaussian states have been studied extensively both
theoretically and experimentally
\cite{Adesso_10,GiordaParis,GAMID,gdexp} but their interplay with
entanglement has not been pinned down so far in terms of the
activation framework. Attempts to devise activation-like protocols
for Gaussian states have been explored \cite{maurosrep}. However,
these differed significantly from the original prescription in
that nonunitary operations were employed between system and
ancillae, so that the entanglement generation was obtained as a
dynamical feature, and conditions i) and ii) were not generally
verified.

Here we consider a general Gaussian activation protocol in
which $U_{A,B}$ are Gaussian unitaries and the CNOT gates are
replaced with a global Gaussian unitary on subsystems $A,B,A', B'$. In Section~\ref{sec_1} we then prove that any such protocol satisfying condition i) will unavoidably violate condition ii), which implies that activation of Gaussian nonclassical correlations by Gaussian operations is impossible. This fact establishes a new no-go theorem for Gaussian quantum information processing, which can be enlisted alongside other well known no-go results such as the the no-distillation theorem, according to which distilling entanglement from Gaussian states by using only Gaussian operations is impossible \cite{nogo}. We then show in Section~\ref{sec_2} how, by using non-Gaussian operations which properly extend the CNOT to infinite dimensions, one can construct the continuous variable counterpart of the activation protocol of \cite{Piani_11}, verifying criteria i) and ii). This allows us to define the negativity of quantumness for Gaussian states and to calculate it for relevant examples in Section~\ref{sec_3}.
This work provides an operational setting to understand and manipulate nonclassical correlations in paradigmatic infinite-dimensional systems. We draw our conclusions in Section~\ref{sec_4}, while some technical derivations (which can be of independent interest) are deferred to the Appendices.

\section{Gaussian no-activation theorem}\label{sec_1}

Gaussian states are quantum states of systems with an infinite-dimensional Hilbert
space (continuous variable systems), e.g.~a collection of harmonic oscillators, which possess a
Gaussian-shaped Wigner function in phase space \cite{ourreview}. $L$ modes are described by a
vector ${\bf r}=\left(x_{1},p_{1},\ldots,x_{L},p_{L}\right)^{T}$
of quadrature operators $x_{j},p_{j}$ satisfying the canonical
commutation rules expressible in terms of elements of the vector
${\bf r}$ as $[r_{j},r_{k}]=i\Omega_{jk}$, $j,k=1,\ldots,L$ with
$\Omega=\oplus_{j=1}^{L}i\sigma_{y}$, where $\sigma_{y}$ is the
Pauli $y$-matrix. An $L$-mode Gaussian state $\rho$ is fully
characterized by a $2L\times 1$ vector $\langle {\bf r}\rangle$ of
the first moments with elements $\langle
r_{i}\rangle=\mbox{Tr}(\rho r_{i})$ and by its $2L\times2L$
covariance matrix (CM) $\gamma$ with elements $\gamma_{ij}=\langle
\Delta r_{i}\Delta r_{j}+\Delta r_{j}\Delta r_{i}\rangle/2$,
$i,j=1,\ldots,L$, where $\Delta r_{i}=r_{i}-\langle r_{i}\rangle$.
Gaussian unitaries are generated by Hamiltonians that are
quadratic in the quadrature operators and they preserve the Gaussian
characteristic of quantum states. An $L$-mode Gaussian unitary $U(S)$ is represented in phase space by a $2L\times2L$ real
symplectic transformation $S$ satisfying the condition $S\Omega
S^{T}=\Omega$, which transforms a CM $\gamma$ to $S\gamma S^{T}$.

Here we are interested in the question of whether an
activation protocol exists satisfying conditions i) and ii) which
would rely solely on Gaussian states and Gaussian unitaries.
We therefore assume the state $\rho_{AB}$ to be a Gaussian state of
$(N+M)$ modes with CM $\gamma_{AB}$ and
the state $\rho_{A'B'}$ of the ancilla to be also a Gaussian state
with CM $\gamma_{A'B'}$. The local unitaries $U_{A,B}$ of the
original discrete protocol are replaced with local Gaussian
unitaries $U_{A}(S_{A})$ and $U_{B}(S_{B})$ represented by the
symplectic matrices $S_{A}$ and $S_{B}$, respectively. Likewise,
the global operation $U_{AA'}^{CNOT}\otimes U_{BB'}^{CNOT}$ on the
whole system $ABA'B'$ is replaced with one global Gaussian unitary
$U(S)$ represented by a symplectic matrix $S$.

Let us recall the definition of a fully classical state
\cite{nolocal,Piani_11,PianiAdesso}. Suppose $\rho_{AB}$ is a
bipartite state containing two subsystems $A$ and $B$ with $N$ and
$M$ modes respectively, and let
$\mathcal{B}_j=\{\ket{{\mathcal{B}_j(\bf n }_{j})}\}$ be a basis
of subsystem $j$, with ${\bf n}_A=(n_{A_1},\ldots,n_{A_N})$, ${\bf
n}_B=({n}_{B_1},\ldots,n_{B_M})$ and $n_{j_i}\in\mathbb{N}_0$. If
there exists a basis $\mathcal{B}$ consisting of the tensor
products of all elements of $\mathcal{B}_A$ with all elements of
$\mathcal{B}_B$, then $\rho_{AB}$ is a classical state if it is
diagonal with respect to $\mathcal{B}$. It has been shown in
\cite{Adesso_10,Rahimi-Keshari_13} that a two-mode Gaussian state
is classical if and only if it is a product state, i.e., its CM is
represented by a direct sum $\gamma_{A}\oplus\gamma_{B}$ of local
CMs $\gamma_{A,B}$. One can prove that this statement remains
valid for the generic case of bipartite $(N+M)$-mode Gaussian
states (see Appendix~\ref{secapp_1} for the proof).
Therefore, all non-product bipartite Gaussian states (including
separable ones) are nonclassical. According to condition i) in any
Gaussian activation protocol with an input Gaussian product state
there must exist local Gaussian unitaries $U_{A,B}$ for which one
gets a separable state $\rho_{ABA'B'}$ across the $AB|A'B'$
splitting at the output of the protocol. We will show however,
that this implies that for {\it all} separable Gaussian states
including nonclassical ones, there exist local Gaussian unitaries
$U_{A,B}$ for which the output state is separable. That is,
condition ii) is not satisfied. Thus, any Gaussian activation
protocol described above cannot meet simultaneously criteria i)
and ii), and hence does not exist.

\begin{figure}[tb]
\includegraphics[width=8.5cm]{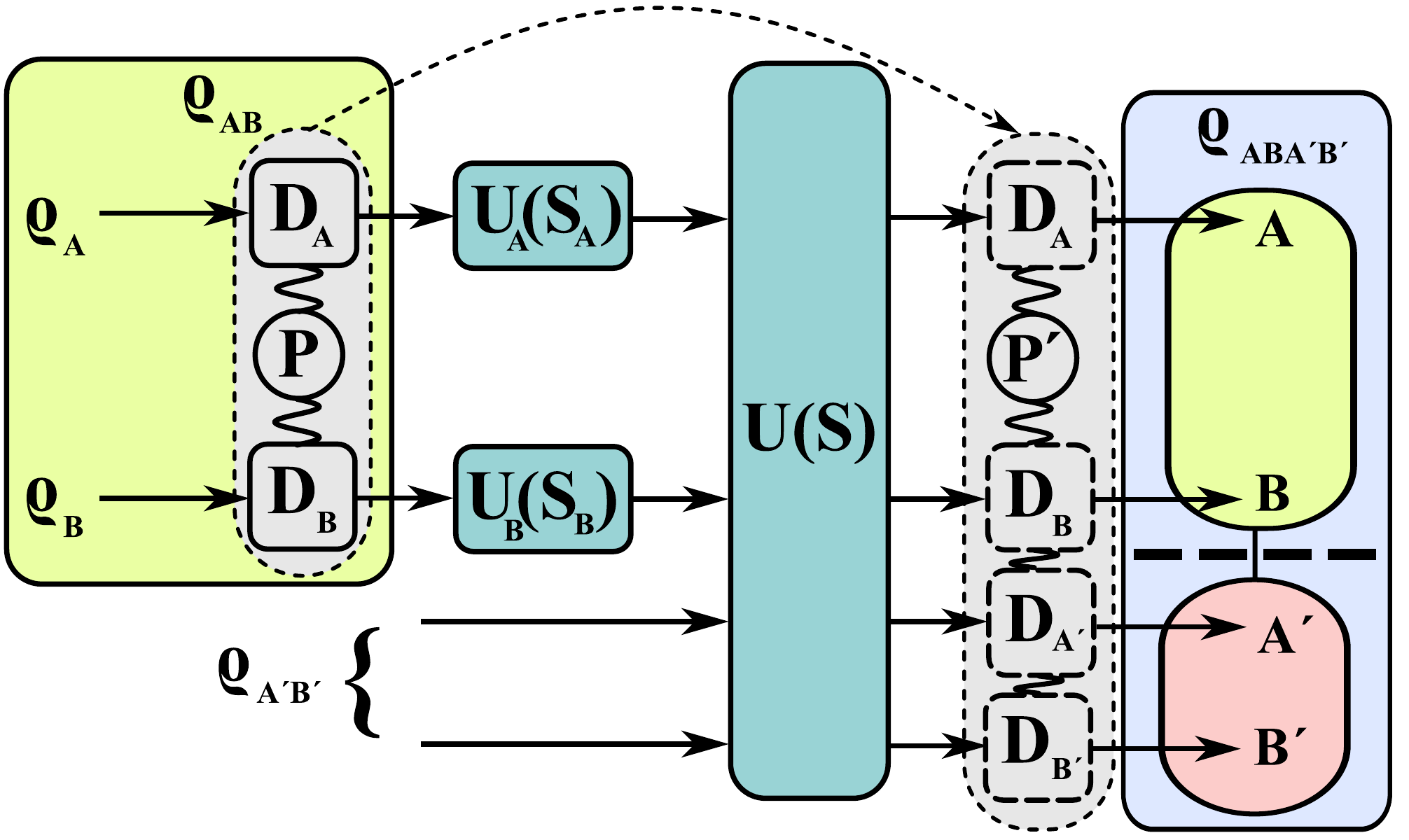}
\caption{(Color online) Pictorial representation of the no-activation theorem.
$\rho_{AB}$ is a separable Gaussian state prepared
from a Gaussian product state $\rho_{A}\otimes\rho_{B}$ by
correlated displacements $D_{A}$ and $D_{B}$ distributed according
to a Gaussian distribution with correlation matrix $P$,
Eq.~(\ref{P}). $U_{A}(S_{A})$ and $U_{B}(S_{B})$ are local
Gaussian unitaries which are adjusted such that without the
displacements $D_{A}$ and $D_{B}$ the activation protocol produces
from the product state $\rho_{A}\otimes\rho_{B}$ an output state
which is separable across the $AB|A'B'$ splitting. As the unitaries
$U_{A}(S_{A})$, $U_{B}(S_{B})$ and $U(S)$ induce a linear
transformation of quadrature operators of the input modes, the
displacements $D_{A}$ and $D_{B}$ can be relocated behind the
global transformation $U(S)$ (dotted arrow). The new
displacements $D_{A},D_{B}, D_{A'}$ and $D_{B'}$,
Eq.~(\ref{displacement}), cannot turn a separable state into an
entangled state and therefore the protocol transforms the
separable state $\rho_{AB}$ into a state which is separable across the
$AB|A'B'$ cut (thick dashed line). See text for details.}
\label{fig1}
\end{figure}

The proof of this no-go theorem is depicted in
Fig.~\ref{fig1}. It follows from the
decomposability of any Gaussian separable state into a product
state and noise \cite{Werner_01}, and the linearity of symplectic
transformations. Namely, for any separable Gaussian state
with CM $\gamma_{AB}$ there exist local CMs $\gamma_{A,B}$ such
that
\begin{equation}\label{P}
P=\gamma_{AB}-\gamma_{A}\oplus\gamma_{B}\geq0.
\end{equation}
In other words, any separable Gaussian state with CM $\gamma_{AB}$
can be prepared from a suitable product state with CM
$\gamma_{A}\oplus\gamma_{B}$ by the addition of noise, represented
by a positive-semidefinite matrix $P$, i.e.,
$\gamma_{AB}=\gamma_{A}\oplus\gamma_{B}+P$. The noise can be
created by displacing the vector of quadratures ${\bf
r}=\left(x_{A_1},p_{A_1},\ldots,x_{A_N},p_{A_N},x_{B_1},p_{B_1},\ldots,x_{B_M},p_{B_M}\right)^{T}$
of the product state as ${\bf r}\rightarrow {\bf r}+{\cal V}{\bf
R}$. Here ${\cal V}$ is a $2(N+M)\times K$ matrix given by the
first $K$ columns of the matrix $V$ bringing the matrix $P$ to the
diagonal form
$V^{T}PV=\mbox{diag}\left(\lambda_1,\lambda_2,\ldots,\lambda_K,0,0,\ldots,0\right)$,
where $\lambda_1,\ldots,\lambda_K$ denote $K\leq2(N+M)$ strictly
positive eigenvalues of the matrix (\ref{P}), and ${\bf
R}=\left(R_1,\ldots,R_K\right)^{T}$ is the vector of classical
displacements uncorrelated with the vector of quadratures ${\bf
r}$ and distributed according to the Gaussian distribution with
zero means and the diagonal correlation matrix
$\mbox{diag}\left(\lambda_1,\lambda_2,\ldots,\lambda_K\right)$.

Let us now consider a separable state with CM $\gamma_{AB}$ at the
input of a Gaussian activation protocol and let
$\gamma_A\oplus\gamma_B$ be a CM of the product state from which
the state can be prepared using the aforementioned algorithm.
Assume that the local symplectic matrices $S_{A,B}$ are chosen
such that the CM $\gamma_{ABA'B'}^{(0)}\equiv S\left(S_{A}\oplus
S_{B}\oplus\openone_{A'B'}\right)\gamma_A\oplus\gamma_B\oplus\gamma_{A'B'}\left(S_{A}^{T}\oplus
S_{B}^{T}\oplus\openone_{A'B'}\right)S^{T}$ of the output state,
where $\gamma_{A'B'}$ is the CM of the state of the ancilla, is
separable across the $AB|A'B'$ splitting. Hence, for the original
separable state with CM $\gamma_{AB}$, the output of the
activation protocol is obtained by displacing the vector of
quadratures ${\bf r}^{(0)}$ for the state with CM
$\gamma_{ABA'B'}^{(0)}$ by
\begin{equation}\label{displacement}
{\bf r}^{(0)}\rightarrow {\bf r}^{(0)}+S\left(\begin{array}{c}
\left(S_A\oplus
S_B\right){\cal V}{\bf R} \\
\mathbb{O} \\
\end{array}\right),
\end{equation}
where $\mathbb{O}$ is a $2T\times 1$ zero vector with $T$ being
the number of modes of the ancilla $A'B'$. However, for a
separable state with CM $\gamma_{ABA'B'}^{(0)}$, where $AB$ is
separable from $A'B'$, the local displacements
(\ref{displacement}) cannot create a state in which the system
$AB$ is entangled with the system $A'B'$. Consequently, for any
separable state (even nonclassical) it is always possible to find
local Gaussian unitaries for which the output is separable, thus
accomplishing the proof of the no-go theorem.

Therefore, Gaussian operations are unable to activate nonclassical correlations of Gaussian separable states into entanglement in the worst-case scenario: assuming condition i) holds, then for any Gaussian separable state there exist local
Gaussian unitaries for which the output of the activation protocol
remains a separable Gaussian state. This indicates that a
non-Gaussian element, like a non-Gaussian global unitary $U$ or a
non-Gaussian state of the ancilla, is
necessary for faithful activation of nonclassical correlations in Gaussian
states. In the following we design such an activation
protocol involving a non-Gaussian CNOT gate in the Fock basis and
an ancillary system in a Gaussian state.

\section{Non-Gaussian activation protocol}\label{sec_2}

The main benefit of the activation protocol is that it allows one
to quantify the amount of nonclassical correlations in a given
quantum state as the potential to create entanglement in the
activation protocol \cite{Piani_11,streltsov,PianiAdesso}. More precisely, if
$E_{AB|A'B'}(\rho_{ABA'B'})$ denotes an entanglement measure
quantifying the amount of entanglement between systems $AB$
and $A'B'$ in a quantum state $\rho_{ABA'B'}$, then we can define
a measure of nonclassical correlations on the input state $\rho_{AB}$ as
\begin{eqnarray}\label{QE}
Q_{E}(\rho_{AB})=\mathop{\mbox{min}}_{U_{A},U_{B}}E_{AB|A'B'}(\rho_{ABA'B'}),
\end{eqnarray}
where the minimization is carried out over all local unitaries $U_{A}$ and $U_{B}$ on subsystems $A$ and $B$.
It has been proven in \cite{PianiAdesso} that  $Q_E(\rho_{AB}) \geq E(\rho_{AB})$, with equality if $\rho_{AB}$ is pure.

From now on we assume that systems $A$ and $B$ each contain one mode. The non-Gaussian activation protocol
is obtained as a direct generalization of the finite-dimensional protocol
\cite{Piani_11}. At the input we allow for generally non-Gaussian
states $\rho_{AB}$ of continuous variable systems, local unitaries $U_A$ and $U_B$, and the global Gaussian unitary $U(S)$ of the preceding protocol is replaced with the
tensor product $V\equiv U_{AA'}^{CNOT}\otimes U_{BB'}^{CNOT}$ of the infinite-dimensional
generalizations of CNOT gates in the Fock basis,
\begin{eqnarray}\label{FockCNOT}
U_{jj'}^{CNOT}|m,n\rangle_{jj'}=|m,m+n\rangle_{jj'},\quad j=A,B,
\end{eqnarray}
where $\ket{m,n}_{jj'}\equiv\ket{m}_{j}\otimes\ket{n}_{j'}$,
$m,n=0,1,\ldots$, and $\ket{k}_l$ is the $k$th Fock state of mode
$l$. We also assume the initial state  $\rho_{A'B'}$ of the
ancilla $A'B'$ to be the vacuum state
$\ket{0}_{A'}\bra{0}\otimes\ket{0}_{B'}\bra{0}$. Hence, the final
output state can be expressed as
\begin{eqnarray}\label{rhoout}
\rho_{ABA'B'}&=&V(\tilde{\rho}\otimes\ket{0}_{A'}\bra{0}\otimes\ket{0}_{B'}\bra{0})V^{\dagger},
\end{eqnarray}
where
\begin{eqnarray}\label{tilderho}
\tilde{\rho}&\equiv& (U_A\otimes
U_B)\rho_{AB}(U^{\dagger}_A\otimes U^{\dagger}_B).
\end{eqnarray}

By following arguments similar to the finite-dimensional case
\cite{Piani_11}, one can show that the non-Gaussian activation
protocol defined above satisfies both criteria i) and ii). For
condition i) we assume that $\rho_{AB}$ is classically correlated
and hence there exist local unitaries $U_A$ and $U_B$ such that
the density matrix $\tilde{\rho}$, Eq.~(\ref{tilderho}), takes the
form
$\tilde{\rho}=\sum_{n,m=0}^{\infty}p_{n,m}\ket{n,m}_{AB}\bra{n,m}$.
Making use of Eqs.~(\ref{FockCNOT}) and (\ref{rhoout}) it then
follows that the  output state of the protocol is the following
convex mixture of product states, $
\rho_{ABA'B'}=\sum_{n,m=0}^{\infty}p_{n,m}\ket{n,m}_{AB}\bra{n,m}\otimes\ket{n,m}_{A'B'}\bra{n,m}
$,
and is thus a separable state across the $AB|A'B'$ splitting
as required.

For the proof of condition ii) we now suppose that the density matrix $\rho_{AB}$ is
nonclassical and show that then the density matrix $\rho_{ABA'B'}$ given in Eq.~(\ref{rhoout}) is entangled across
the $AB|A'B'$ cut for all local unitaries $U_A$ and $U_B$. To prove the presence of entanglement
in $\rho_{ABA'B'}$ we will use the negativity $\mathcal{N}$ defined in \cite{Vidal_02} as
\begin{equation}\label{N1}
\mathcal{N}(\rho_{ABA'B'})=\frac{1}{2}(\|\rho^{T_{AB}}_{ABA'B'}\|_1-1).
\end{equation}
Here $\|.\|_1$ denotes the trace norm, $\rho^{T_{AB}}_{ABA'B'}$ is the partial transpose \cite{Peres_96} of the state $\rho_{ABA'B'}$
with respect to subsystem $AB$, and a strictly positive value of negativity implies that the state $\rho_{ABA'B'}$ is (distillable) entangled with respect to the
$AB|A'B'$ splitting. The specific feature of the present activation protocol is that the output state $\rho_{ABA'B'}$ is a so-called maximally
correlated state and therefore, following results in \cite{Piani_11,Takafumi_13}, the output negativity can be expressed as
\begin{equation}\label{N2}
\mathcal{N}(\rho_{ABA'B'})=\frac{1}{2}\sum_{\mathbf{m\neq
n}=\mathbf{0}}^{\infty}|\tilde{\rho}_{\mathbf{m},\mathbf{n}}| = \frac{1}{2}\left(\sum_{\mathbf{m,n}=\mathbf{0}}^{\infty}|\tilde{\rho}_{\mathbf{m},\mathbf{n}}|-1\right),
\end{equation}
where $\mathbf{m}=(m_{1},m_{2})$, $\mathbf{n}=(n_{1},n_{2})$ and $\tilde{\rho}_{\mathbf{m},\mathbf{n}}=_{AB}\langle m_{1}m_{2}|\tilde{\rho}|n_{1}n_{2}\rangle_{AB}$ are elements of the density matrix $\tilde{\rho}$, Eq.~(\ref{tilderho}), in the Fock basis.

Since our input state $\rho_{AB}$ is nonclassical, the state $\tilde{\rho}$ is also nonclassical for any choice of unitaries $U_A$ and $U_B$.
Thus, there must be at least one non-zero off-diagonal element $\tilde{\rho}_{\mathbf{m},\mathbf{n}}$ for every choice of $U_A$ and $U_B$. Hence, Eq.~(\ref{N2})
implies $\mathcal{N}(\rho_{ABA'B'})>0$ and the output state $\rho_{ABA'B'}$ is entangled for any nonclassical input state. This completes the proof of our non-Gaussian activation protocol.

\section{Examples}\label{sec_3}

The optimization in Eq.~(\ref{QE}) is generally carried out over all local
unitary operations $U_{A}$ and $U_{B}$, including non-Gaussian ones,
which is not a tractable task. Here we consider input Gaussian states with CM in standard form \cite{Simon_00}, and consider the non-optimized output entanglement $E_{AB|A'B'}(\rho_{ABA'B'})$  obtained when the local unitaries $U_{A,B}$ are selected to be identity matrices. Therefore, the state
 (\ref{tilderho}) remains a Gaussian state in standard form with
the following CM
\begin{eqnarray}\label{tildegamma}
\tilde{\gamma}=\left(\begin{array}{cc}
A & C \\
C^T & B\\
\end{array}\right),
\end{eqnarray}
where $A=\mbox{diag}(a,a)$, $B=\mbox{diag}(b,b)$ and $C=\mbox{diag}(c_1,c_2)$ are diagonal matrices.
In what follows we determine the non-optimized
quantity for some classes of two-mode Gaussian states by considering the negativity (\ref{N1}) as an entanglement measure $E$, and using Eq.~(\ref{N2}).
The corresponding measure of nonclassical correlations $Q_{\mathcal{N}}(\rho_{AB})$
is called the negativity of quantumness \cite{Piani_11} accordingly. Although our choice of local unitaries $U_{A,B}$
gives in general an upper bound on  $Q_{\mathcal{N}}$, we find that it coincides with the
true measure on pure states, leading us to conjecture that our choice is optimal for calculating the negativity of quantumness of all two-mode Gaussian
states in standard form. Verifying this conjecture numerically is beyond the scope of this work.

\subsection{Pure states} 
\begin{figure}[tb]
\includegraphics[width=8.5cm]{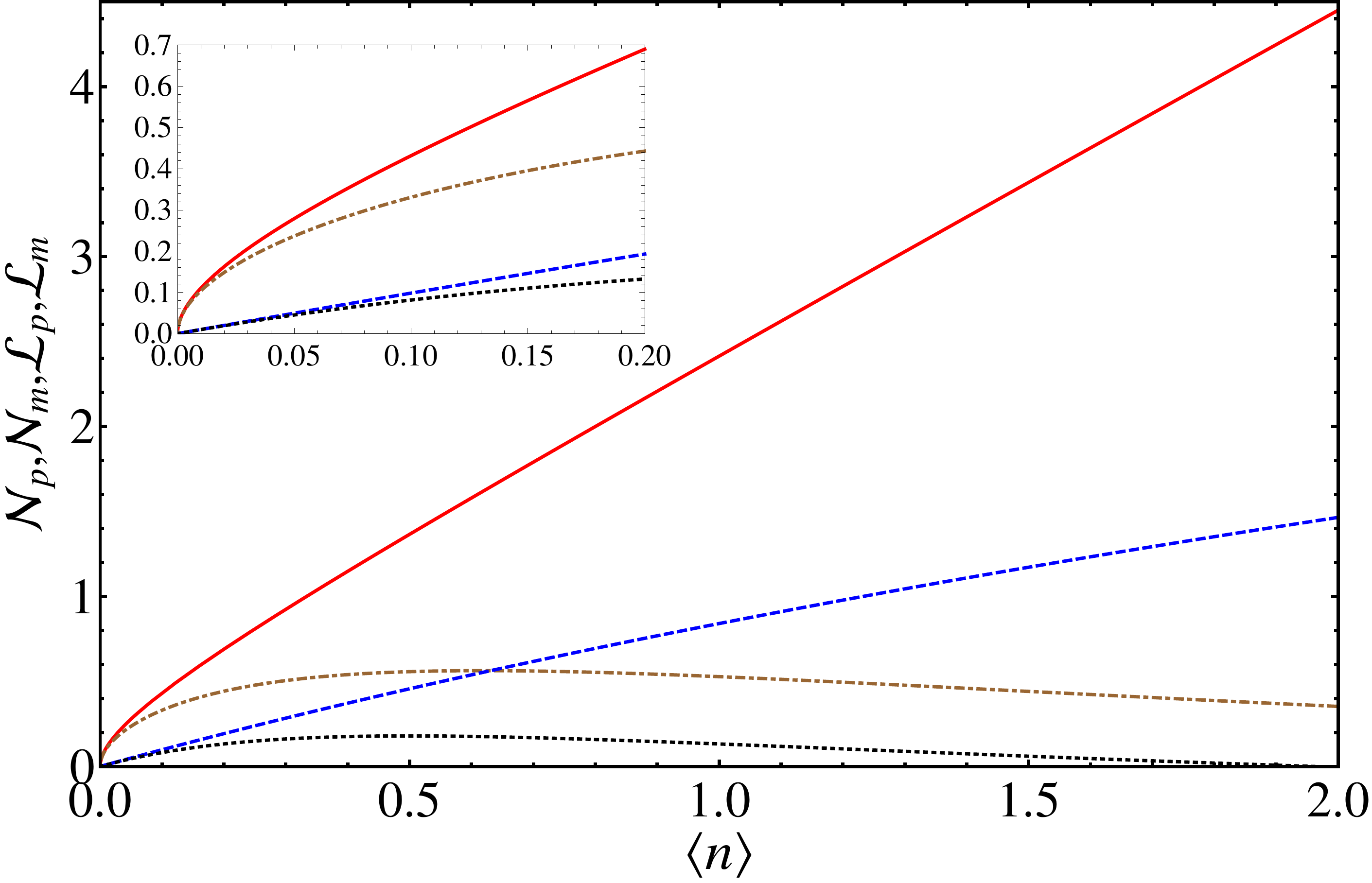}
\caption{(Color online) Negativity of quantumness $\mathcal{N}_{\rm p}$ [Eq.~(\ref{Np})] (solid red
line) and its lower bound $\mathcal{L}_{\rm p}$ [Eq.~(\ref{Lp})] (dash-dotted brown line)
for pure squeezed vacuum states, plotted as a function of the local mean number of thermal photons
$\langle n\rangle=\sinh^2(r)$. Upper bound on the negativity of quantumness $\mathcal{N}_{\rm m}$ [Eq.~(\ref{Nm})] (dashed blue line)
and its lower bound $\mathcal{L}_{\rm m}$ [Eq.~(\ref{Lm})] (dotted black line) for separable mixed
states obtained as unbiased mixtures of coherent states, plotted as a function of the local mean number of thermal photons $\langle n\rangle=\sigma^2$. The inset shows a close-up for $\langle n \rangle \ll 1$, where the lower bounds become tight.}\label{fig2}
\end{figure}

A closed form of the output negativity can be found for pure two-mode Gaussian
states. The density matrix $\tilde{\rho}$ amounts to that of the two-mode squeezed vacuum
state, with $\tilde{\rho}_{\mathbf{m},\mathbf{n}}=[1-\tanh^2(r)](\tanh r)^{m_1+n_1}\delta_{m_1,m_2}\delta_{n_1,n_2}$,
where $r\geq0$ is the squeezing parameter. Hence, by a direct substitution into Eq.~(\ref{N2})
we get
\begin{equation}\label{Np}
\mathcal{N}_{\rm p}=\mbox{$\frac12(e^{2r}-1)$}.
\end{equation}
Consequently, as the output negativity $\mathcal{N}(\rho_{ABA'B'})$ is equal to the negativity
of the input state $\rho_{AB}$, it coincides with the true optimized
negativity of quantumness $Q_{\mathcal{N}}(\rho_{AB})$ \cite{PianiAdesso}, and our
choice of local unitaries is thus optimal for pure states.
The negativity (\ref{Np}) is depicted by a solid red line in
Fig.~\ref{fig2}.

\subsection{Unbiased mixtures of coherent states}
These Gaussian states are of the form
\begin{equation}\label{mixture}
\rho_{AB}=\int_{\mathbb{C}} P(\alpha)|\alpha\rangle_{A}\langle\alpha|\otimes|\alpha\rangle_{B}\langle\alpha|d^{2}\alpha,
\end{equation}
and can be prepared by splitting a thermal state with
mean number of thermal photons $2\sigma^2$ on a balanced beam
splitter. Here $\alpha\in\mathbb{C}$,
$P(\alpha)=\mbox{exp}(-|\alpha|^2/\sigma^{2})/(\pi\sigma^2)$ and
$d^{2}\alpha=d(\mbox{Re}\alpha)d(\mbox{Im}\alpha)$. The states are
already in standard form with a CM (\ref{tildegamma}) specified by
$a=b=\sigma^2+1/2$ and $c_1=c_2=\sigma^2$. Making use of
the components of a coherent state in Fock basis $\langle
m|\alpha\rangle=\mbox{exp}(-|\alpha|^2/2)\alpha^{m}/\sqrt{m!}$ we
get the following matrix elements of the state (\ref{mixture}),
\begin{equation}\label{mixtureFock}
\tilde{\rho}_{\mathbf{m},\mathbf{n}}=\frac{(m_1+m_2)!}{\sqrt{m_1!m_2!n_1!n_2!}}\frac{\delta_{m_1+m_2,
n_1+n_2}}{s(m_1+m_2)},
\end{equation}
where $s(j)=\sigma^2\left(1/\sigma^2+2\right)^{j+1}$. By
substitution of the latter expression into Eq.~(\ref{N2}) we get
after some algebra
\begin{equation}\label{Nm}
\mathcal{N}_{\rm m}=\frac{1}{2}\left\{\sum_{M=0}^{\infty}\frac{1}{s(M)}\left[\sum_{J=0}^{M}\sqrt{{M \choose J}}\right]^2-1\right\}.
\end{equation}
The negativity (\ref{Nm}) is depicted by a dashed blue line in
Fig.~\ref{fig2}, and is generally smaller than the one of pure states calculated in (\ref{Np}). Both classes of Gaussian states have a nonzero negativity of quantumness which increases with $\langle n \rangle >0$; this is in agreement with earlier studies of nonclassical correlations based on entropic measures of quantum discord \cite{Adesso_10,GiordaParis}.

\subsection{Standard-form two-mode Gaussian states}
In general we need the Fock
basis elements $\tilde{\rho}_{\mathbf{m},\mathbf{n}}$ for an
arbitrary two-mode Gaussian state with zero first moments. Combining the
results of Refs.~\cite{Dodonov_84,Dodonov_94,Fiurasek_review01} we
can express them as
\begin{equation}\label{rhoHermite}
\tilde{\rho}_{\mathbf{m},\mathbf{n}}=\frac{H_{m_1,m_2,n_1,n_2}^{(R)}(0)}{\sqrt{\text{det}\left(\tilde{\gamma}+\frac{1}{2}\openone\right)}\sqrt{m_1!m_2!n_1!n_2!}},
\end{equation}
where $\tilde{\gamma}$ is the CM of the state, $\openone$ is the
$4\times 4$ identity matrix, and $H_{m_1,m_2,n_1,n_2}^{(R)}(0)$ is
the four-dimensional Hermite polynomial \cite{Bateman_53} at the
origin; see Appendix~\ref{secapp_2} for a complete derivation of
Eq.~(\ref{rhoHermite}). Here
\begin{equation}\label{RR}
R=WO\left[\left(\tilde{\gamma}+\frac{1}{2}\openone\right)^{-1}-\openone\right]O^{\dagger}V
\end{equation}
is the symmetric matrix defining the polynomial, where
\begin{equation}
W=\left(\begin{array}{cccc}
1 & 0 & 0 & 0\\
0 & 0 & 1 & 0\\
0 & 1 & 0 & 0\\
0 & 0 & 0 & 1\end{array}\right),\quad V=\left(\begin{array}{cccc}
0 & 0 & 1 & 0\\
1 & 0 & 0 & 0\\
0 & 0 & 0 & 1\\
0 & 1 & 0 & 0\end{array}\right),
\end{equation}
and
\begin{eqnarray}\label{OO}
O=\bigoplus_{j=1}^{2}\frac{1}{\sqrt{2}}\left(\begin{array}{cc}
1 & i\\
1 & -i \end{array}\right).
\end{eqnarray}
For the standard-form CM $\tilde{\gamma}$, Eq.~(\ref{tildegamma}),
we get in particular
\begin{equation}\label{Rst}
R=\left(\begin{array}{cc}
R_{1}-R_{2} & R_{1}+R_{2}-\openone_2 \\
R_{1}+R_{2}-\openone_2 & R_{1}-R_{2}\\
\end{array}\right)
\end{equation}
with $\openone_2$ being the $2\times 2$ identity matrix,
\begin{equation}\label{Rj}
R_{j}=\displaystyle\frac{1}{2d_{j}}\left(\begin{array}{cc}
b+\frac{1}{2} & -c_{j} \\
-c_{j} & a+\frac{1}{2}\\
\end{array}\right),
\end{equation}
and $d_{j}=(a+1/2)(b+1/2)-c_{j}^2$  ($j=1,2$). One can then
evaluate the negativity (\ref{N2}) by performing a numerical
summation of the absolute values of the elements
(\ref{rhoHermite}). The higher-order Hermite polynomials can be
calculated from the lower-order ones by using e.g.~the recurrence
formula derived in Appendix~\ref{secapp_2}.

We remark that the compact expression in equation (\ref{rhoHermite}) is of independent interest and can be useful for the characterization of hybrid information processing involving conversion between continuous and discrete variable entanglement \cite{hybridrev}, or particularly for studies of Bell nonlocality of arbitrary two-mode Gaussian states by means of dichotomic pseudospin measurements \cite{chen}, whose expectation value can be conveniently evaluated at the Fock space level.

In the context of the present paper, apart from the utility for numerical evaluation of the output negativity (\ref{N2}), equation (\ref{rhoHermite}) also enables us to derive a simple analytical lower bound on the output negativity. The bound results from the following chain
of inequalities
\begin{eqnarray}\label{inequalities}
\sum_{\mathbf{m,n}=\mathbf{0}}^{\infty}|\tilde{\rho}_{\mathbf{m},\mathbf{n}}|&=&
\sum_{\mathbf{m,n}=\mathbf{0}}^{\infty}\frac{|H_{m_1,m_2,n_1,n_2}^{(R)}(0)|}
{\sqrt{\text{det}\left(\tilde{\gamma}+\frac{1}{2}\openone\right)}\sqrt{m_1!m_2!n_1!n_2!}}\nonumber\\
&\geq&\sum_{\mathbf{m,n}=\mathbf{0}}^{\infty}\frac{|H_{m_1,m_2,n_1,n_2}^{(R)}(0)|}
{\sqrt{\text{det}\left(\tilde{\gamma}+\frac{1}{2}\openone\right)}{m_1!m_2!n_1!n_2!}}\nonumber\\
&\geq&\vline\sum_{\mathbf{m,n}=\mathbf{0}}^{\infty}\frac{H_{m_1,m_2,n_1,n_2}^{(R)}(0)}
{\sqrt{\text{det}\left(\tilde{\gamma}+\frac{1}{2}\openone\right)}{m_1!m_2!n_1!n_2!}}\vline\nonumber\\
&=&\frac{e^{-\frac{1}{2}\sum_{i,j=1}^{4}R_{ij}}}
{\sqrt{\mbox{det}\left(\tilde{\gamma}+\frac{1}{2}\openone\right)}},
\end{eqnarray}
where the first inequality follows from the inequality $1/\sqrt{n!}\geq1/{n!}$ which holds for any $n\geq0$, the second
inequality is a consequence of the triangular inequality for absolute values, and the last equation follows
from the expression for the generating function of the four-dimensional Hermite polynomials at the origin \cite{Bateman_53},
\begin{equation}\label{generatingfunction}
e^{-\frac{1}{2}h^T R h}=\sum_{\mathbf{m,n}=\mathbf{0}}^{\infty}{\frac{{\alpha_1^*}^{m_1}{\alpha_2^*}^{m_2}\alpha_1^{n_1}\alpha_2^{n_2}}{m_1!m_2!n_1!n_2!}}
H_{m_1,m_2,n_1,n_2}^{(R)}(0),
\end{equation}
where $h=(\alpha_1^*,\alpha_2^*,\alpha_1,\alpha_2)^T$ and $R$ is the matrix (\ref{RR}).
A comparison between the right-hand side (RHS) of the previous equation and the expression of the
Husimi $Q$-quasiprobability distribution
$\Phi_{\mathcal{A}}(\alpha_1,\alpha_2)=\bra{\alpha_1\alpha_2}\tilde{\rho}\ket{\alpha_1\alpha_2}/\pi^2$ in the Fock
basis further yields
\begin{equation}\label{PhiAR}
\frac{e^{-\frac{1}{2}h^T R h}}{\sqrt{\mbox{det}\left(\tilde{\gamma}+\frac{1}{2}\openone\right)}}=\pi^2e^{|\alpha_1|^2+|\alpha_{2}|^2}\Phi_{\mathcal{A}}(\alpha_1,\alpha_2)
\end{equation}
as can be easily seen from the results of Appendix~\ref{secapp_2}.
Therefore, the last expression in the chain of inequalities (\ref{inequalities}) can be written in the following compact form
\begin{equation}\label{PhiA11}
\frac{e^{-\frac{1}{2}\sum_{i,j=1}^{4}R_{ij}}}
{\sqrt{\mbox{det}\left(\tilde{\gamma}+\frac{1}{2}\openone\right)}}=\left(\pi e\right)^2\Phi_{\mathcal{A}}(1,1).
\end{equation}
Now, making use of the inequalities (\ref{inequalities}) and equality (\ref{PhiA11}) one finds that the sum in (\ref{N2}) is lower-bounded as
\begin{equation}\label{sumbound}
\sum_{\mathbf{m,n}=\mathbf{0}}^{\infty}|\tilde{\rho}_{\mathbf{m},\mathbf{n}}|\geq\left(\pi e\right)^2\Phi_{\mathcal{A}}(1,1),
\end{equation}
which finally gives the following bound on the output negativity (\ref{N2})
\begin{equation}\label{L}
\mathcal{N}(\rho_{ABA'B'})\geq\frac{1}{2}\left[\left(\pi e\right)^2\Phi_{\mathcal{A}}(1,1)-1\right].
\end{equation}

The bound (\ref{L}) can be evaluated for any zero-mean two-mode Gaussian state with CM $\tilde{\gamma}$ by
calculating the matrix (\ref{RR}) and substituting it into the formula (\ref{PhiA11}). To test
the tightness of the bound we calculate it for the previous examples of pure states and mixtures of coherent states, and compare the obtained lower bounds with the exact values of the negativities (\ref{Np}) and (\ref{Nm}), respectively.
The CM $\tilde{\gamma}$ is in the standard form (\ref{tildegamma}) in both cases and therefore one can evaluate
easily the matrix (\ref{RR}) using Eqs.~(\ref{Rst}) and (\ref{Rj}) which gives, after substitution into Eq.~(\ref{PhiA11}),
\begin{equation}\label{PhiA11p}
\left(\pi e\right)^2\Phi_{\mathcal{A}}^{\rm p}(1,1)=\frac{e^{2\tanh r}}
{\cosh^2(r)}
\end{equation}
for pure states, and
\begin{equation}\label{PhiA11m}
\left(\pi e\right)^2\Phi_{\mathcal{A}}^{\rm m}(1,1)=\frac{e^{\frac{4\sigma^{2}}{2\sigma^{2}+1}}}{2\sigma^{2}+1}
\end{equation}
for unbiased mixtures of coherent states. The corresponding negativities then satisfy
\begin{equation}\label{Lp}
\mathcal{N}_{\rm p}\geq\frac{1}{2}\left[\frac{e^{2\tanh r}}
{\cosh^2(r)}-1\right]\equiv\mathcal{L}_{\rm p}
\end{equation}
and
\begin{equation}\label{Lm}
\mathcal{N}_{\rm m}\geq\frac{1}{2}\left(\frac{e^{\frac{4\sigma^{2}}{2\sigma^{2}+1}}}{2\sigma^{2}+1}-1\right)\equiv\mathcal{L}_{\rm m}.
\end{equation}
The bounds $\mathcal{L}_{\rm p}$ and $\mathcal{L}_{\rm m}$ as well as the negativities $\mathcal{N}_{\rm p}$, Eq.~(\ref{Np}), and $\mathcal{N}_{\rm m}$, Eq.~(\ref{Nm}), are depicted in Fig.~\ref{fig2}. The figure shows that both bounds are tight in the region of small $\langle n \rangle$ (see the inset), which also proves that Eq.~(\ref{Nm}) amounts to the exact value of the negativity of quantumness for mixtures of coherent states with small mean number of thermal photons in each mode. Both lower bounds are then shown to increase with increasing $\langle n\rangle$ and the gap between the bounds ${\cal L}_{\rm p, m}$ and the numerically evaluated values of the output negativities ${\cal N}_{\rm p, m}$ gets larger. Further analysis reveals however that the lower bounds $\mathcal{L}_{\rm p}$ and $\mathcal{L}_{\rm m}$ are nonmonotonic for larger $\langle n \rangle$; they both attain a maximum at
$\langle n\rangle \approx 0.62$ and $\langle n\rangle \approx 0.52$, respectively, and then both monotonically decrease for larger values of $\langle n\rangle$; eventually, both lower bounds become trivial as they enter the region of negative values, namely $\mathcal{L}_{\rm p}<0$ for
$\langle n\rangle\gtrsim5.26$ and $\mathcal{L}_{\rm m}<0$ for $\langle n\rangle\gtrsim 1.97$.

As a final remark, note that the sum in negativity (\ref{N2}) just amounts to the so-called $\ell_1$-norm of the density matrix $\tilde{\rho}$ \cite{Takafumi_13}, i.e.,
$\sum_{\mathbf{m,n}=\mathbf{0}}^{\infty}|\tilde{\rho}_{\mathbf{m},\mathbf{n}}|=\|\tilde{\rho}\|_{\ell_1}$.
The results of the present Section thus also describe how to calculate numerically the $\ell_1$-norm for an arbitrary two-mode Gaussian state with zero means and the inequality (\ref{sumbound}) gives a simple analytical lower bound $\|\tilde{\rho}\|_{\ell_1}\geq\left(\pi e\right)^2\Phi_{\mathcal{A}}(1,1)$ on such a norm.
\section{Conclusions}\label{sec_4}

We have shown that a protocol capable of activating
nonclassical correlations in bipartite Gaussian states based
solely on Gaussian operations cannot exist.
We have also
constructed a non-Gaussian activation protocol and we have
investigated quantitatively its performance using the negativity
of quantumness as a figure of merit. Our analysis suggests that
optimal performance of the protocol is achieved if the input
Gaussian state is in the standard form. Restricting to the local
Gaussian unitaries the conjecture can be proved or disproved with
the help of Eq.~(\ref{rhoHermite}) by numerical minimization of
the negativity (\ref{N2}) with respect to the unitaries, which is
left for further research.

We believe that our results
will stimulate further exploration of the negativity of
quantumness and its interplay with other nonclassicality
indicators \cite{Adesso_10,GAMID} in the context of Gaussian
states.
\acknowledgments

L.~M. acknowledges the Project No. P205/12/0694 of GA\v{C}R and the European Social Fund and MSMT
under project No. EE2.3.20.0060. D.~M. acknowledges the support of
the Operational Program Education for Competitiveness Project No.
CZ.1.07/2.3.00/20.0060 co-financed by the European Social Fund and
Czech Ministry of Education. G.~A. acknowledges the Brazilian
agency CAPES [Pesquisador Visitante Especial-Grant No.~108/2012]
and the Foundational Questions Institute [Grant No.
FQXi-RFP3-1317]. G.~A. would also like to thank M. Barbieri and M. Piani for discussions.

\appendix

\section{Classically correlated bipartite Gaussian states are product
states}\label{secapp_1}

This section is dedicated to the proof that a bipartite Gaussian
state $\rho_{AB}$ of an $N$-mode subsystem $A$ and an $M$-mode
subsystem $B$ is classically correlated across the $A|B$ splitting if
and only if it is a product state $\rho_{A}\otimes\rho_{B}$.

The proof of the ``only if'' part is trivial because any product
state is diagonal in the product of eigenbases of local states.

The ``if'' part can be proved using the necessary and sufficient
condition for zero quantum discord \cite{Rahimi-Keshari_13}.
Quantum discord $D_{B}(\rho_{AB})$ of a quantum state $\rho_{AB}$
with a measurement on subsystem $B$ is zero if an only if the state
can be expressed as \cite{Datta_08}
\begin{equation}\label{QCstate}
\rho_{AB}=\sum_{i}p_{i}\rho_{A}^{(i)}\otimes|i\rangle_{B}\langle
i|,\quad 0\leq p_{i}\leq 1,
\end{equation}
where $\{|i\rangle_{B}\}$ is an orthonormal basis of subsystem $B$. The
zero-discord criterion \cite{Rahimi-Keshari_13} then says that a
quantum state $\rho_{AB}$ can be expressed in the form
(\ref{QCstate}) if and only if for an informationally complete
positive operator valued measurement (IC-POVM) on subsystem $A$,
the conditional states $\rho_{B|k}$ of subsystem $B$ corresponding
to the measurement outcomes $k$, mutually commute, i.e.,
\begin{equation}\label{criterion}
[\rho_{B|k},\rho_{B|k'}]=0,\quad \mbox{for all $k$ and $k'$}.
\end{equation}

We consider a Gaussian state $\rho_{AB}$ with zero means and
covariance matrix (CM) $\gamma_{AB}$. Modes
$A_{1},A_{2},\ldots,A_{N}$ comprising the subsystem $A$ are
subject to a Gaussian measurement characterized by a CM
$\gamma_{\rm m}$ and a vector of measurement outcomes
$k=(x_{A_{1}},p_{A_{1}},\ldots,x_{A_{N}},p_{A_{N}})^T\in\mathbb{R}_{2N}$.
If a measurement outcome $k$ occurs then the state $\rho_{AB}$
collapses into the $M$-mode state $\rho_{B|k}$ of subsystem $B$
with CM $\sigma$ and vector of first moments $d_{k}$ of the form
\cite{Giedke_02}
\begin{eqnarray}
\sigma&=&B-C^{T}\frac{1}{A+\gamma_{\rm m}}C,\label{sigma}\\
d_{k}&=&C^{T}\frac{1}{A+\gamma_{\rm m}}k,\label{dk}
\end{eqnarray}
where $A,B$ and $C$ are blocks of the CM $\gamma_{AB}$ expressed
with respect to the $A|B$ splitting,
\begin{equation}\label{gammaABblock}
\gamma_{AB}=\left(\begin{array}{cc}
A & C\\
C^T & B \end{array}\right).
\end{equation}

As in Ref.~\cite{Rahimi-Keshari_13} we will now express criterion (\ref{criterion}) in terms of the characteristic
function. For this purpose we will first use the fact that an
$M$-mode quantum state $\rho_{j}$ can be expressed as
\cite{Giedke_02}
\begin{equation}\label{rhoj}
\rho_{j}=\frac{1}{(2\pi)^{M}}\int_{\mathbb{R}_{2M}}C_{j}(\xi)W^{\dag}(\xi)d\xi,
\end{equation}
where $C_{j}(\xi)$ is the characteristic function of the state
$\rho_{j}$ and $W(\xi)=\mbox{exp}(-i\xi^{T}\mathbf{r})$ is the
displacement operator with
$\xi=(\xi_{x_{1}},\xi_{p_{1}},\ldots,\xi_{x_{M}},\xi_{p_{M}})^{T}\in\mathbb{R}_{2M}$
and $\mathbf{r}=(x_{1},p_{1},\ldots,x_{M},p_{M})^{T}$ is the
vector of quadratures. Due to the validity of the relation
$\mbox{Tr}\left[W^{\dag}(\xi')W(\xi)\right]=(2\pi)^{M}\delta(\xi-\xi')$
we get from Eq.~(\ref{rhoj}) immediately the following expression
for the characteristic function of the state $\rho_{j}$:
\begin{equation}\label{Cj}
C_{j}(\xi)=\mbox{Tr}\left[\rho_{j}W(\xi)\right].
\end{equation}
Making use of Eq. (\ref{rhoj}) we can express the commutator on the left-hand side (LHS) of Eq.~(\ref{criterion}) as
\begin{widetext}
\begin{equation}\label{commutator}
[\rho_{B|k},\rho_{B|k'}]=\frac{1}{(2\pi)^{2M}}\int\int_{\mathbb{R}_{2M}}C_{k}(\xi)C_{k'}(\xi')
\left(e^{-\frac{i}{2}\xi^{T}\Omega\xi'}-e^{\frac{i}{2}\xi^{T}\Omega\xi'}\right)W^{\dag}(\xi+\xi')d\xi
d\xi',
\end{equation}
\end{widetext}
where $C_{k}(\xi)$ and $C_{k'}(\xi')$ are the characteristic
functions of the states $\rho_{B|k}$ and $\rho_{B|k'}$,
respectively, and where we have used the relation
\begin{equation}\label{WdagW}
W^{\dag}(\xi')W(\xi)=e^{\frac{i}{2}\xi'^{T}\Omega\xi}W(\xi-\xi'),
\end{equation}
with
\begin{equation}\label{Omega}
\Omega=\bigoplus_{i=1}^{M}\left(\begin{array}{cc}
0 & 1\\
-1 & 0 \end{array}\right).
\end{equation}
From Eqs.~(\ref{sigma}) and (\ref{dk}) it follows that the
Gaussian states $\rho_{B|k}$ and $\rho_{B|k'}$ possess the same CM
$\sigma$ and first moments $d_{k}$ and $d_{k'}$, respectively, and
therefore their characteristic functions read
\begin{equation}
C_{k}(\xi)=e^{-\frac{1}{2}\xi^{T}\sigma\xi-i\xi^{T}d_{k}},\quad
C_{k'}(\xi')=e^{-\frac{1}{2}\xi'^{T}\sigma\xi'-i\xi'^{T}d_{k'}}.
\end{equation}

Equation (\ref{commutator}) allows us to calculate the
characteristic function of the commutator
$[\rho_{B|k},\rho_{B|k'}]$ given by
\begin{equation}\label{Ckkprimed}
C_{kk'}(\xi)=\mbox{Tr}\left\{[\rho_{B|k},\rho_{B|k'}]W(\xi)\right\}.
\end{equation}
By inserting the RHS of the commutator from Eq.~(\ref{commutator})
into Eq.~(\ref{Ckkprimed}), using Eq.~(\ref{WdagW}) and carrying
out the integration, we arrive at the characteristic function
(\ref{Ckkprimed}) in the form
\begin{widetext}
\begin{equation}\label{Ckkprimedfinal}
C_{kk'}(\xi)=2\frac{e^{-\frac{1}{4}\xi^{T}\left(\sigma+\frac{1}{4}\Omega^{T}\sigma^{-1}\Omega\right)\xi-\frac{1}{4}\left(d_{k}-d_{k'}\right)^{T}\sigma^{-1}\left(d_{k}-d_{k'}\right)
-\frac{i}{2}\xi^{T}(d_{k}+d_{k'})}}{2^{M}\sqrt{\mbox{det}\sigma}}\sinh\left[\frac{1}{4}\left(d_{k'}-d_{k}\right)^{T}\sigma^{-1}\Omega\xi\right].
\end{equation}
\end{widetext}
From Eq.~(\ref{Ckkprimed}) and the formula
\begin{equation}\label{commutatorviaC}
[\rho_{B|k},\rho_{B|k'}]=\frac{1}{(2\pi)^{M}}\int_{\mathbb{R}_{2M}}C_{kk'}(\xi)W^{\dag}(\xi)d\xi
\end{equation}
it follows that $[\rho_{B|k},\rho_{B|k'}]=0$ if and only if
$C_{kk'}(\xi)=0$ for all $\xi$. Assuming that the CM $\sigma$ in
Eq.~(\ref{sigma}) has finite second moments and the measurement
outcomes $k$ and $k'$ and hence also the displacements $d_{k}$ and
$d_{k'}$ defined by Eq.~(\ref{dk}) are finite, the condition
$C_{kk'}(\xi)=0$ for all $\xi$ is equivalent to the condition
\begin{equation}\label{condition1}
\left(d_{k'}-d_{k}\right)^{T}\sigma^{-1}\Omega\xi=0,\quad \mbox{for all $\xi$},
\end{equation}
which can be rewritten using Eq.~(\ref{dk}) as
\begin{equation}\label{condition2}
\left(k'-k\right)^{T}\frac{1}{A+\gamma_{\rm
m}}C\sigma^{-1}\Omega\xi=0.
\end{equation}

Previous results allow us to rephrase the zero-discord criterion
of Ref.~\cite{Rahimi-Keshari_13} for bipartite Gaussian states and
Gaussian IC-POVMs as follows. An $N+M$-mode Gaussian state
$\rho_{AB}$ can be expressed in the form (\ref{QCstate}) if and
only if the condition (\ref{condition2}) is satisfied for all
$k,k'$, where $k$ and $k'$ are measurement outcomes of
an Gaussian IC-POVM on subsystem $A$ characterized by the CM
$\gamma_{\rm m}$. Condition
(\ref{condition2}) is satisfied for all $k,k'$ ($k\neq k'$) if
and only if the matrix
\begin{equation}\label{condition3}
\frac{1}{A+\gamma_{\rm m}}C\sigma^{-1}\Omega=0.
\end{equation}
Consider now the heterodyne measurement which is an example of a
Gaussian IC-POVM \cite{D'Ariano_04}. Then $\gamma_{\rm
m}=(1/2)\openone$, the matrix $\frac{1}{A+\gamma_{\rm m}}$ is
invertible and therefore condition (\ref{condition3}) is
equivalent with the equation $C\sigma^{-1}\Omega=0$. As both the
matrices $\Omega$ and $\sigma^{-1}$ are also invertible the latter
condition is equivalent with the condition $C=0$. For the
heterodyne detection the condition (\ref{condition3}) is thus
equivalent with the vanishment of the off-diagonal block $C$ given
in Eq.~(\ref{gammaABblock}), which carries intermodal
correlations. This means in other words, that a bipartite
$(N+M)$-mode Gaussian state can be expressed in the form
(\ref{QCstate}) if and only if it is a product state, i.e.,
$\rho_{AB}=\rho_{A}\otimes\rho_{B}$.

Let us now move to the necessary and sufficient condition for a
bipartite $(N+M)$-mode Gaussian state to be a classically correlated
state. A quantum state $\rho_{AB}$ is classically correlated if
and only if $D_{A}(\rho_{AB})=D_{B}(\rho_{AB})=0$
\cite{Bylicka_10}, where $D_{A}(\rho_{AB})$ is the discord of
$\rho_{AB}$ for measurement on subsystem $A$. A quantum state
$\rho_{AB}$ is therefore classically correlated if and only if it can be
expressed simultaneously in the form (\ref{QCstate}) and in the
form
\begin{equation}\label{CQstate}
\rho_{AB}=\sum_{i}p_{i}|i\rangle_{A}\langle
i|\otimes\rho_{B}^{(i)}.
\end{equation}
According to the criterion given in \cite{Rahimi-Keshari_13} a quantum
state $\rho_{AB}$ can be expressed in the form (\ref{CQstate}) if
and only if for an IC-POVM on subsystem $B$ the conditional states
$\rho_{A|k}$ of subsystem $A$ corresponding to the measurement
outcomes $k$ mutually commute, i.e.,
\begin{equation}\label{commutator2}
[\rho_{A|k},\rho_{A|k'}]=0,\quad \mbox{for all $k$ and $k'$}.
\end{equation}
Like in the previous case we can express the latter condition in
terms of a characteristic function. We can proceed exactly along
the same lines as in the case of the commutator (\ref{commutator})
with the only difference that now we consider measurement on the
$M$-mode subsystem $B$. Consequently, the formulas which we get
for the present case of the commutator (\ref{commutator2}) are
obtained from the formulas derived in the context of commutator
(\ref{commutator}) by the replacements $A\leftrightarrow B$,
$C\leftrightarrow C^{T}$ of the blocks of the matrix $\gamma_{AB}$
and by the replacement $M\rightarrow N$. Thus we find that the
commutator (\ref{commutator2}) vanishes if and only if $C^{T}=0$.
Therefore, the condition $C=0$ is necessary and sufficient for an
$(N+M)$-mode Gaussian state to be classical, which concludes
our proof.

\section{Matrix elements of a Gaussian state in Fock basis in terms of Hermite
polynomials}\label{secapp_2}

Our aim is to express the elements of a density matrix of a
Gaussian state $\rho$ of two modes $A$ and $B$ in the Fock basis.
Here and in what follows we assume that the state has all first
moments equal to zero. The present derivation combines the results
obtained in Refs.~\cite{Dodonov_84,Dodonov_94,Fiurasek_review01}.
Firstly we express the elements of the density matrix in the basis
of coherent states as
\begin{eqnarray}\label{cohelements}
&&e^{|\alpha_1|^2+|\alpha_2|^2}\bra{\alpha_1\alpha_2}\rho\ket{\alpha_1\alpha_2} \\ && \quad  =\sum_{m_1,m_2,n_1,n_2=0}^{\infty}
\frac{{\alpha_1^*}^{m_1}{\alpha_2^*}^{m_2}\alpha_1^{n_1}\alpha_2^{n_2}}{\sqrt{m_1!m_2!n_1!n_2!}}\bra{m_1m_2}\rho\ket{n_1n_2}, \nonumber
\end{eqnarray}
where we have used the expression of the components of a coherent
state $|\alpha\rangle$ in the Fock basis
\begin{equation}\label{alpham}
\langle
m|\alpha\rangle=e^{-\frac{|\alpha|^2}{2}}\frac{\alpha^{m}}{\sqrt{m!}}.
\end{equation}
The matrix element on the LHS of
Eq.~(\ref{cohelements}) can be further expressed as
\begin{equation}\label{LHS}
\bra{\alpha_1\alpha_2}\rho\ket{\alpha_1\alpha_2}=\pi^2\Phi_{\mathcal{A}}(\alpha_1,\alpha_2),
\end{equation}
where
\begin{equation}\label{PhiA}
\Phi_{\mathcal{A}}(\alpha_1,\alpha_2)=\frac{1}{\pi^2\sqrt{\text{det}\gamma_{\mathcal{A}}^{(c)}}}e^{-\frac{1}{2}\alpha^{\dagger}\left[{\gamma_{\mathcal{A}}}^{(c)}\right]^{-1}\alpha}
\end{equation}
is the Husimi $Q$-quasiprobability distribution of the
Gaussian state $\rho$ \cite{Perina_91}. Here,
$\alpha=(\alpha_1,\alpha_1^*,\alpha_2,\alpha_2^*)^T$ and
$\gamma_{\mathcal{A}}^{(c)}$ is the  complex
CM corresponding to antinormal ordering of the canonical operators.
Substituting now from Eq.~(\ref{LHS}) into the LHS of
Eq.~(\ref{cohelements}) and making use of Eq.~(\ref{PhiA}) we
arrive at the following equality
\begin{eqnarray}\label{Focksum}
&&\frac{1}{\sqrt{\text{det}\gamma_{\mathcal{A}}^{(c)}}}e^{-\frac{1}{2}\alpha^{\dagger}\left\{\left[\gamma_{\mathcal{A}}^{(c)}\right]^{-1}-\openone\right\}\alpha}\\
&& \quad =\sum_{m_1,m_2,n_1,n_2=0}^{\infty}\frac{{\alpha_1^*}^{m_1}{\alpha_2^*}^{m_2}\alpha_1^{n_1}\alpha_2^{n_2}}{\sqrt{m_1!m_2!n_1!n_2!}}\bra{m_1m_2}\rho\ket{n_1n_2}. \nonumber
\end{eqnarray}

The LHS of the latter equation can be expressed in terms
of the multi-dimensional Hermite polynomials \cite{Bateman_53}.
Specifically, the generating function of the four-dimensional
Hermite polynomials is
\begin{equation}\label{generatingfunction}
e^{-\frac{1}{2}h^T R
h+h^TRx}=\sum_{m_1,m_2,n_1,n_2=0}^{\infty}{\frac{{\alpha_1^*}^{m_1}{\alpha_2^*}^{m_2}\alpha_1^{n_1}\alpha_2^{n_2}}{m_1!m_2!n_1!n_2!}}
H_{m_1,m_2,n_1,n_2}^{(R)}(x),
\end{equation}
where $h=(\alpha_1^*,\alpha_2^*,\alpha_1,\alpha_2)^T$,
$x=(x_1,x_2,x_3,x_4)^T$, and $R$ is a symmetric matrix of order
four. The LHS of Eq.~(\ref{Focksum}) then can be rewritten in terms of the LHS of Eq.~(\ref{generatingfunction}) as follows. The
complex CM $\gamma_{\mathcal{A}}^{(c)}$ can be expressed as
\begin{equation}\label{gammaAc}
\gamma_{\mathcal{A}}^{(c)}=O\left(\gamma+\frac{1}{2}\openone\right)O^{\dagger},
\end{equation}
where $\openone$ is the $4\times 4$ identity matrix,
\begin{eqnarray}\label{O}
O=\bigoplus_{j=1}^{2}\frac{1}{\sqrt{2}}\left(\begin{array}{cc}
1 & i\\
1 & -i \end{array}\right)
\end{eqnarray}
is a $4\times 4$ unitary matrix, and $\gamma$ is the standard real
symmetrically ordered CM of the state $\rho$, with elements
$\gamma_{ij}=\langle r_{i}r_{j}+r_{j}r_{i}\rangle/2$,
$i,j=1,\ldots,4$, where $r_{i}$ is the $i$-th component of the
vector of quadratures ${\bf
r}=\left(x_{A},p_{A},x_{B},p_{B}\right)^{T}$. Hence we get
\begin{equation}\label{gammaAc}
\left[\gamma_{\mathcal{A}}^{(c)}\right]^{-1}-\openone=O\left[\left(\gamma+\frac{1}{2}\openone\right)^{-1}-\openone\right]O^{\dagger}.
\end{equation}
Furthermore, we can write
\begin{equation}\label{alpha}
\alpha=Vh,\quad \alpha^{\dag}=h^{T}W,
\end{equation}
where
\begin{equation*}
V=\left(\begin{array}{cccc}
0 & 0 & 1 & 0\\
1 & 0 & 0 & 0\\
0 & 0 & 0 & 1\\
0 & 1 & 0 & 0\end{array}\right),\quad W=\left(\begin{array}{cccc}
1 & 0 & 0 & 0\\
0 & 0 & 1 & 0\\
0 & 1 & 0 & 0\\
0 & 0 & 0 & 1\end{array}\right).
\end{equation*}
Consequently,
\begin{equation*}
\alpha^{\dagger}\left\{\left[\gamma_{\mathcal{A}}^{(c)}\right]^{-1}-\openone\right\}\alpha=h^T
Rh,
\end{equation*}
where
\begin{equation}\label{R}
R=WO\left[\left(\gamma+\frac{1}{2}\openone\right)^{-1}-\openone\right]O^{\dagger}V.
\end{equation}
As $(WO)^{T}=O^{\dag}V$ and the CM $\gamma$ is symmetric, one finds immediately that
$R^{T}=R$ and therefore $R$ is symmetric as required. Making use of
Eqs.~(\ref{Focksum}) and (\ref{generatingfunction}) we get
\begin{eqnarray*}
&& \frac{1}{\sqrt{\text{det}\gamma_{\mathcal{A}}^{(c)}}}\sum_{m_1,m_2,n_1,n_2=0}^{\infty}\frac{{\alpha_1^*}^{m_1}{\alpha_2^*}^{m_2}\alpha_1^{n_1}\alpha_2^{n_2}}{m_1!m_2!n_1!n_2!} H_{m_1,m_2,n_1,n_2}^{(R)}(0)\\ && \quad =
\sum_{m_1,m_2,n_1,n_2=0}^{\infty}\frac{{\alpha_1^*}^{m_1}{\alpha_2^*}^{m_2}\alpha_1^{n_1}\alpha_2^{n_2}}{\sqrt{m_1!m_2!n_1!n_2!}}\bra{m_1m_2}\rho\ket{n_1n_2},
\end{eqnarray*}
where the matrix $R$ defining the Hermite polynomial
$H_{m_1,m_2,n_1,n_2}^{(R)}$ is given in Eq.~(\ref{R}). By equating each term in the summation we are left with the elements
of the density matrix $\rho$ in the Fock basis,
\begin{equation}\label{rhoFock}
\bra{m_1m_2}\rho\ket{n_1n_2}=\frac{H_{m_1,m_2,n_1,n_2}^{(R)}(0)}{\sqrt{\text{det}\gamma_{\mathcal{A}}^{(c)}}\sqrt{m_1!m_2!n_1!n_2!}},
\end{equation}
where
\begin{equation}\label{detgammaAc}
\mbox{det}\gamma_{\mathcal{A}}^{(c)}=\mbox{det}\left(\gamma+\frac{1}{2}\openone\right).
\end{equation}
Equation (\ref{rhoFock}) allows us to calculate any element of
a density matrix in the Fock basis for an arbitrary two-mode Gaussian
state with zero first moments.

To calculate matrix (\ref{R}) it is convenient to
express the CM $\gamma$ in the block form
\begin{equation}\label{gammablock}
\gamma=\left(\begin{array}{cc}
A & C\\
C^T & B \end{array}\right).
\end{equation}
This allows us to express the inverse matrix
$(\gamma+\openone/2)^{-1}$, appearing in Eq.~(\ref{R}), in block
form using the following blockwise inversion formula \cite{Horn_85},
\begin{eqnarray}\label{blockwise}
\!\!\!\!\!\! &\!\!\!\!\!\! &\!\!\!\!\!\!\left(\begin{array}{cc}
A & C\\
C^{T} & B\\
\end{array}\right)^{-1}
\\
\!\!\!\!\!\!  &\!\!\!\!\!\! &\!\!\!\!\!\! \  =\left(\begin{array}{cc}
\left(A-CB^{-1}C^{T}\right)^{-1} & A^{-1}C\left(C^{T}A^{-1}C-B\right)^{-1}\\
\left(C^{T}A^{-1}C-B\right)^{-1}C^{T}A^{-1} & \left(B-C^{T}A^{-1}C\right)^{-1}\\
\end{array}\right). \nonumber
\end{eqnarray}
\subsection{Recurrence relations}

Higher-order Hermite polynomials can be calculated from lower-order polynomials using a recurrence relation. It is
derived from the generating function (\ref{generatingfunction}),
where we set $x=0$. By deriving both sides of the equation
(\ref{generatingfunction}) with respect to the $i$-th element of
the vector $h=(\alpha_1^*,\alpha_2^*,\alpha_1,\alpha_2)^T$,
substituting the RHS of Eq.~(\ref{generatingfunction}) for the
exponential function $\mbox{exp}\left(-h^T R h/2\right)$ appearing
on the LHS of the obtained expression and equating each term in
the summation, we arrive at the following recurrence relation
\begin{equation}
H_{\mu+e_i}^{(R)}(0)=-\sum_{j=1}^4r_{ij}\mu_jH_{\mu-e_j}^{(R)}(0),
\end{equation}
where $H_{\mu}^{(R)}(0)$ is the four-dimensional Hermite
polynomial at the origin with multi-index $\mu=(m_1,m_2,n_1,n_2)$. The coefficients
$r_{ij}$ correspond to the $(i,j)$-th element of the matrix $R$,
Eq.~(\ref{R}), and $e_i$ is the $i$-th canonical basis vector
with 1 in the $i$-th component and zeros everywhere else. Here,
any Hermite polynomial with a negative index is zero, i.e.
$H_{\mu}(0)=0$ for all $\mu$ with $\mu_i<0$ for some $i$. Every
Hermite polynomial at the origin can be found from the latter
recurrence formula and by using the first few cases,
\begin{align}
H_{0,0,0,0}^{(R)}(0)&= 1,\\
H_{e_i}^{(R)}(0)&=0,\\
H_{e_i+e_j}^{(R)}(0)&=-r_{ij},\\
H_{e_i+e_j+e_k}^{(R)}(0)&=0,\\
H_{1,1,1,1}^{(R)}(0)&=r_{12}r_{34}+r_{23}r_{41}+r_{13}r_{24},
\end{align}
with $i\neq j \neq k$. These can be derived by a direct
calculation from the expression
\begin{eqnarray}
H_{\mu}^{(R)}(x)&=&(-1)^{\sum_{i=1}^{2}
n_i+m_i}\exp\left(\frac{1}{2}x^TRx\right) \\
&\times& \frac{\partial^{\sum_{i=1}^2
n_i+m_i}}{\partial {x_1^{m_1}}\partial {x_2^{m_2}}\partial
{{x}_3^{n_1}}\partial
{{x}_4^{n_2}}}\exp\left(-\frac{1}{2}x^TRx\right),
\nonumber \end{eqnarray}
found in \cite{Dodonov_84}. Note that it is sufficient to calculate only the polynomials
where the parity of the multi-index $\mu$ is even. When the parity of the multi-index $\mu$ is odd, i.e.
$P(\mu)=m_1+m_2+n_1+n_2=2\ell+1$, where $\ell \in \mathbb{N}_0$,
then $H_{\mu}^{(R)}(0)=0$.


\end{document}